\documentclass[twoside]{dis09}
\usepackage[latin1]{inputenc}
\usepackage{subfig}
\usepackage[dvips]{graphicx,epsfig,color}
\usepackage{wrapfig,rotating}
\usepackage{amssymb,amsmath,array}

\begin{document}
\title{Diffraction and Vector Mesons Working Group Summary}

\author{Markus Diehl$^1$, Paul Laycock$^2$ and Christophe Royon$^3$
\vspace{.3cm}\\
1- Deutsches Elektronen-Synchroton DESY \\
22603 Hamburg, Germany
\vspace{.1cm}\\
2- University of Liverpool, Dept.\ of High Energy Physics \\
Oliver Lodge Laboratory, Liverpool L69 7ZE, UK
\vspace{.1cm}\\
3- CEA/IRFU/Service de Physique des Particules \\
91191 Gif-sur-Yvette CEDEX, France
}

\maketitle

\begin{abstract}
We summarise the talks presented in the working group on diffraction and
vector mesons at the DIS 2009 workshop.
\end{abstract}

\section{Experiment}

\subsection{Exclusive vector meson production and DVCS}

Exclusive vector meson production provides an ideal experimental
testing ground for QCD, as the experimental signature is clean and a
variety of theoretical calculations is available.  A comprehensive
review of exclusive vector meson production data at HERA was reported
in \cite{Levy-talk,Marage-talk}.  Particular attention was drawn to
the excellent agreement between the H1 and ZEUS collaborations on
$\rho$ meson electro-production data, while the apparent discrepancies
in the $\phi$ data may be understood at least qualitatively.  The
universality of vector meson production was addressed in particular in
\cite{Levy-talk}, where the conclusions are yet to be finalised, but
certainly the data prefer a universal scale like $\frac{1}{4}
(Q^2+M^2)$ rather than $Q^2$ and follow a power-law in the scaling
variable.

The spin density matrix element analysis of these data exhibits a
wealth of information, as reported in \cite{Marage-talk}.  One
hitherto unnoticed observation is that the double-flip amplitude may
be non-zero, which may be the first hint for a nonzero gluon
transversity distribution \cite{Kivel:2001qw}.  The data exhibit a
clear breaking of $s$-channel helicity conservation, which remains
however a good approximation at low $t$.  Generally, the kinematic
dependences of both the cross sections and the spin density matrix
elements are at least qualitatively understood in QCD.  The data can
be described using models based on generalised parton distributions
(GPDs) or on the dipole approach, with gross features being described
but differences in the detail.

\begin{figure}[h]
\begin{center}
\subfloat[\label{Fig:dvcs}]{\includegraphics[width=0.40\columnwidth]{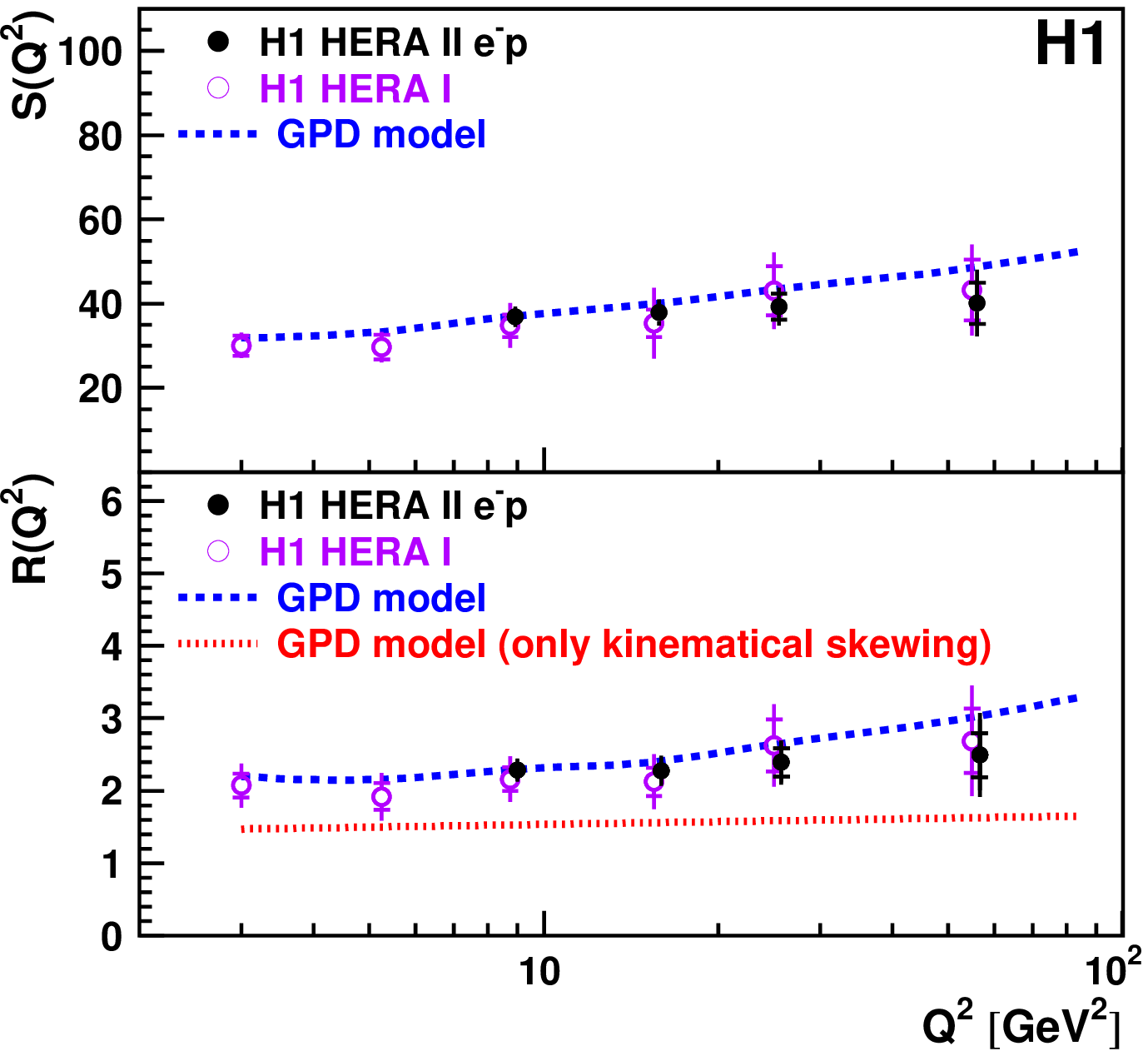}}
\subfloat[\label{Fig:hight}]{\includegraphics[width=0.45\columnwidth]{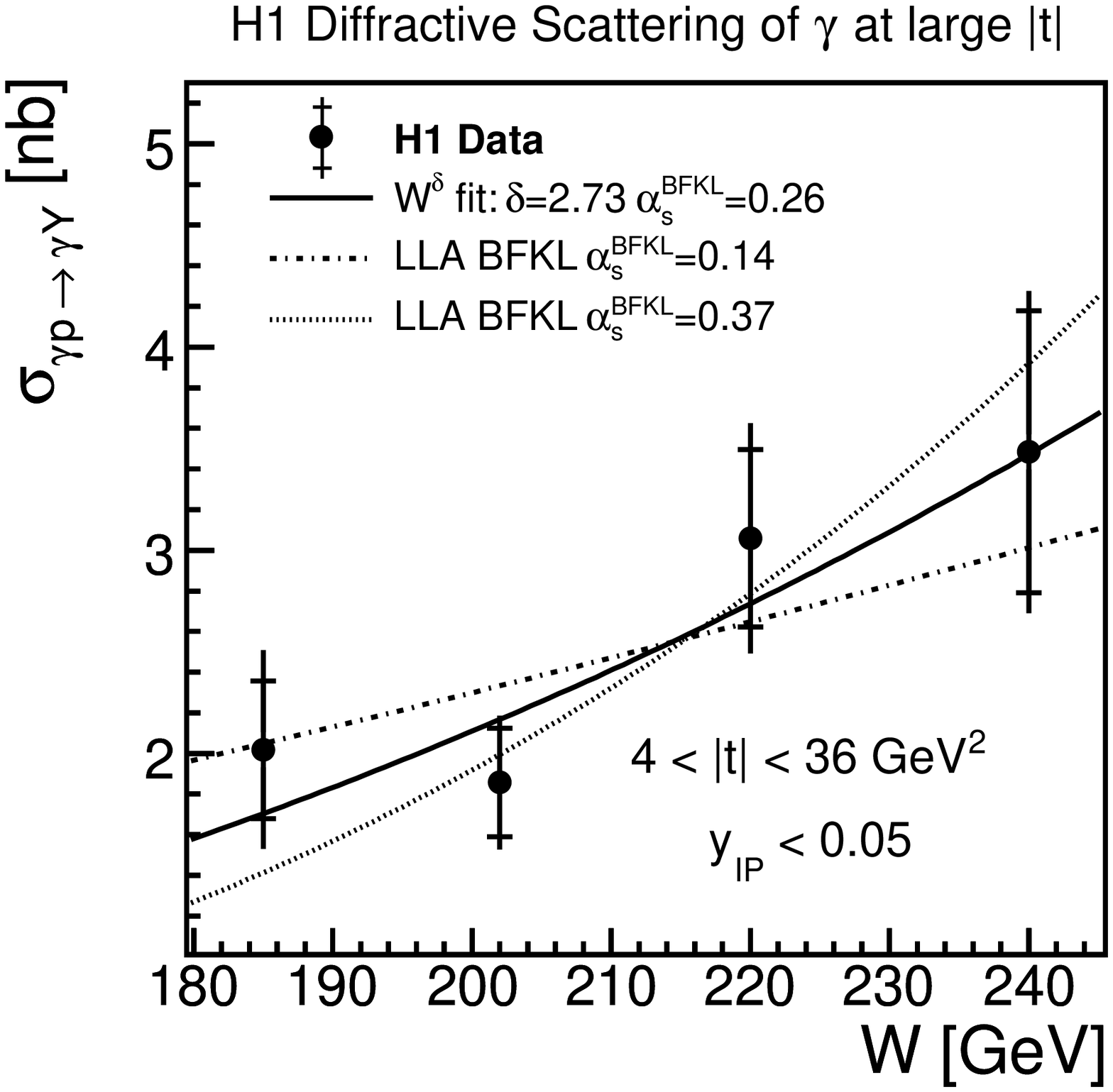}}
\caption{(a) The $Q^2$ dependence of quantities sensitive to GPDs
  (see text) and (b) the $W$ dependence of the high-$t$ photon
  cross section.}
\end{center}
\end{figure}

Deeply virtual Compton scattering is a process with sensitivity to
correlations of partons in the proton and has sensitivity to GPDs, as
shown in \cite{Aaron:2007cz}.  Fig.  $\ref{Fig:dvcs}$ shows (top) the
$Q^2$ dependence of a dimensionless variable $S$ related to the
amplitude for the process with the $t$-dependence removed; (bottom)
the $Q^2$ dependence of a variable $R$ related to the ratio of GPD to
PDF.  The data are precise enough to discriminate between GPD models
and favour a full GPD model rather than one with only kinematical
skewing.

Measured at H1, the exclusive production of photons at high momentum
transfer $t$ at the proton vertex was presented in \cite{Hreus-talk}.
These data allow comparison of the experimental results with
predictions based on a BFKL approach; thanks to the final state
photon, these predictions do not suffer from uncertainty on the final
state vector meson wave function.  The gross features of the kinematic
dependences of the data, i.e. the $W$ and $t$ dependences, are well
reproduced by the model.  The $W$ dependence of the high-$t$ photon
cross section is shown in Fig.~$\ref{Fig:hight}$; this is certainly
one of the hardest diffractive processes yet measured, with an
exponent $\delta = 2.73 \pm 1.02
\,\text{(stat.)}^{+0.56}_{-0.78}\,\text{(sys.)}$.  This is consistent
with the model predictions, although the precision of the data is
limited experimentally by the small lever-arm in $W$.

The Pomeron trajectory in $\rho$ photo-production is one of the
cleanest processes to measure experimentally and is assumed to be one
of the cleanest to interpret theoretically, as it approaches the bare
Pomeron.  A new global fit to high precision H1 data at low $t$
combined with ZEUS and Omega data at higher $t$ was reported in
\cite{List-talk}.  The global result, dominated by the low $t$ data,
shows clear evidence of a non-linear trajectory; the high precision
low $t$ data in particular show a marked departure from a straight
line.  Although this is not at all in disagreement with theory, it is
an interesting result and in contradiction with the usual assumption
of a linear trajectory.

\begin{figure}[ht]
\begin{center}
\includegraphics[width=0.4\columnwidth]{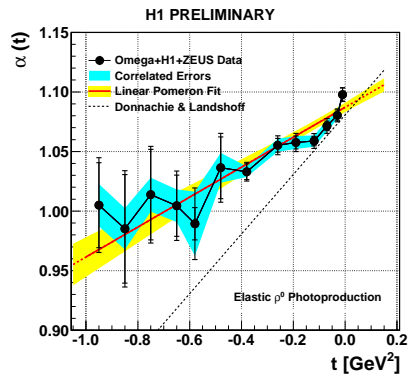}
\caption{The Pomeron trajectory extracted from a global fit to $\rho$
  photoproduction data.}
\label{Fig:rhotraj}
\end{center}
\end{figure}

Elastic $\rho$ production has also been studied by the COMPASS
collaboration, where the asymmetry in angle between the target and
hadron production planes is measured \cite{Jegou-talk}.  One key
feature of the experimental setup is that three targets are used,
allowing the asymmetry to be defined in such a way that the
experimental effects completely cancel.  The measured asymmetry is
consistent with zero, in agreement with a previous measurement by the
HERMES collaboration and also consistent with theoretical expectations
based on GPD models.

A future measurement of DVCS at medium $x$ is planned at COMPASS, with
high precision measurements of the $t$-slope of the process and the
beam charge and spin asymmetry foreseen \cite{Burtin-talk}.  The
promising results from a 2-day test run from 2008 show a detector
performance consistent with the expectations of the technical design
report.

The PHENIX collaboration have made a first measurement of the
exclusive production of J/$\Psi$ mesons in ultra-peripheral Au-Au
collisions at RHIC, reported in \cite{Csanad-talk}.  This difficult
measurement required a very good understanding of the significant
irreducible continuum background.  The measured cross section agrees
with theoretical predictions very well and, although the statistics at
RHIC are limited, the agreement with expectation bodes very well for
the LHC where this process will be measured in a new energy regime.

\subsection{Leading baryon production}

Leading neutrons have been studied by the PHENIX collaboration and the
left-right production asymmetry measured, as reported in
\cite{Togawa-talk}.  A significant asymmetry is observed and both this
asymmetry and the cross section measured differentially in
longitudinal momentum fraction is consistent with models based on the
pion exchange mechanism.  Similar observations were reported by the
ZEUS \cite{Bruni-talk} and H1 \cite{Dodonov-talk} collaborations for
large values of the longitudinal momentum of the neutron.

Leading protons have been studied by both the ZEUS \cite{Bruni-talk}
and H1 \cite{Kapishin-talk} collaborations, where the latter reported
a new measurement on the full HERA II data set.  The conclusions,
especially on the proton vertex factorisation assumption, remain the
same as for previous analyses, i.e. proton vertex factorisation holds
within the large normalisation uncertainties which dominate the
results.  It is hoped that the Very Forward Proton Spectrometer (VFPS)
of H1, installed for the HERA II period, will soon be able to shed
more light on these issues with greater precision, thanks to its much
larger acceptance.

\subsection{Forward detectors at the LHC}
\begin{figure}
\begin{center}
\epsfig{file=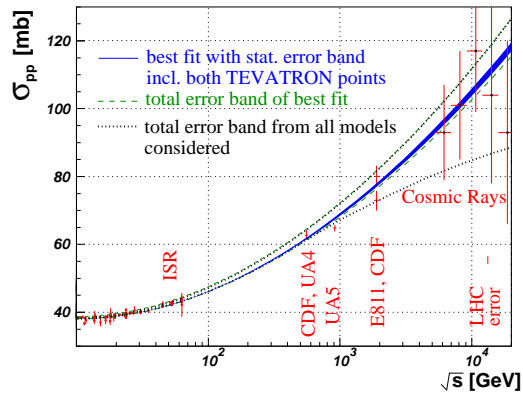,width=7cm}
\caption{Measurement of the total $pp$ cross section.}
\label{total}
\end{center}
\end{figure}

Measurements of the total cross section and luminosity are foreseen in
the ATLAS-ALFA~\cite{Pinfold-talk} and TOTEM~\cite{Niewiadomski-talk}
experiments. Roman pots are installed at 147 and 220 m in TOTEM and at
240 m in ATLAS, and additional forward detectors in TOTEM called T1,
T2 cover the rapidity regions $3.1<|\eta|<4.7$ and $5.3<|\eta|<6.5$.
The measurement of the total cross section to be performed by the
TOTEM collaboration~\cite{Niewiadomski-talk} is shown in
Fig.~\ref{total}.  We notice that there is a large uncertainty on
predictions of the total cross section at the LHC energy due in
particular to the discrepancy between the Tevatron measurements (CDF,
E710 and E811).  The measurement of TOTEM will be of special interest
to solve that ambiguity.  The TOTEM experiment will also allow to
perform early measurements of single diffraction and double pomeron
exchange using the horizontal detectors.

The ATLAS collaboration prefers to measure the elastic scattering in
the Coulomb region~\cite{Pinfold-talk}, typically at very low $t$ ($|t|
\sim 6.5 \times 10^{-4}$ GeV$^2$). When $t$ is close to 0, the $t$
dependence of the elastic cross section reads:
\begin{eqnarray}
\left. \frac{dN}{dt} \right|_{t \rightarrow 0} = L \pi\, \left|\,
     \frac{2 \alpha_{\text{em}}}{|t|} -
\frac{\sigma_{\text{tot}}}{4 \pi} (i+\rho)\, e^{-b|t|/2} \,\right|^2 \;.
\end{eqnarray}
{}From a fit to the data in the Coulomb region, it is possible to
determine directly the total cross section $\sigma_{tot}$, the $\rho$
and $b$ parameters as well as the absolute luminosity $L$. This
measurement requires to go down to $t \sim 6.5 \times 10^{-4}$ GeV$^2$, or
$\theta \sim 3.5~\mu$rad (to reach the kinematical domain where the
strong amplitude equals the electromagnetic one).
Once the absolute luminosity and the total cross section are known
using these methods, the relative luminosity measurement as a function
of time will be performed in ATLAS using the LUCID detector
(Luminosity measurement Using Cerenkov Integrating
Detectors)~\cite{Pinfold-talk}.

\subsubsection{Hard diffraction at the LHC}
The LHC with a centre-of-mass energy of 14 TeV will allow us to access
a completely new kinematical domain in diffraction.  The study of hard
diffraction at the LHC will be performed using the rapidity gap method
at low luminosity (without pileup) and the forward detectors at
higher luminosity. With 10 pb$^{-1}$, about 300 single diffractive
dijet events are expected while about 100 single diffractive $W$
events can be measured with 100 pb$^{-1}$. The correlation between the
multiplicities in low and high $|\eta|$ (respectively $3.0<|\eta|<4.0$
and $ 4.0<|\eta|<5.0$) allows the isolation of diffractive
events~\cite{Obertino-talk}. The single diffractive measurements will
determine the survival probability at LHC energies, as well
as the rate and kinematic dependences of exclusive $\Upsilon$
photoproduction.

At higher luminosities, the CMS and ATLAS collaborations project to
install forward detectors at 220-240 and 420
m~\cite{Pinfold-talk,Piotrzkowski-talk}. These detectors will allow the
study of hard diffractive production of heavy objects such as the
Higgs boson or the study of anomalous couplings between $\gamma$ and
$W$~\cite{Piotrzkowski-talk}.  Two kinds of detectors namely 3D
Si and timing detectors, will be hosted in movable beam pipes located
at 220-240 and 420 m.  At the LHC, up to 35 interactions occur at the
same bunch crossing; in order to determine which interaction a proton
tagged in one of these detectors belongs to a timing resolution of
better than 10-15 ps is needed.  Lab results reported in
\cite{Piotrzkowski-talk} have demonstrated that this is indeed
possible.

\subsection{Diffractive results from the Tevatron}
The CDF experiment at the Tevatron showed many new diffractive
results~\cite{Mesropian-talk}.  The main central CDF detectors
(tracker and calorimeter) cover respectively the regions $|\eta| <2.0$
and $|\eta| <3.6$.  Beam shower counters and miniplug calorimeters
extend the rapidity coverage between $5.4 < |\eta| < 7.4$ and $3.5 <
|\eta| < 5.1$, respectively.  Roman pot detectors were installed only
on the $\bar{p}$ side and allow the detection of antiprotons with
$0.02 < \xi < 0.1$ ($\xi$ is the momentum fraction of the proton
carried away by the pomeron). The extended coverage in rapidity using
the miniplug calorimeters was especially important to reconstruct
$\xi$ more precisely.

\subsubsection{Diffractive $W$ and $Z$ production}
The CDF collaboration measured the diffractive production of $W$
and $Z$ bosons, which probes the quark content of the pomeron. The
fraction of $W$ events tagged in the Roman pot detectors is $0.97 \pm
0.05 \,\text{(stat.)} \pm 0.11 \,\text{(syst.)}$ in the kinematical domain
with $0.03 < \xi < 0.1$ and $|t|<1$ GeV$^2$, compatible with run I
results, while the fraction of $Z$ events is $0.85 \pm 0.20
\,\text{(stat.)} \pm 0.11 \,\text{(syst.)}$ 
in the same kinematical domain.

\subsubsection{Jet-gap-jet and Mueller-Navelet jets}
The miniplug calorimeters of the CDF collaboration allow a detailed
study of rapidity gaps between jets (double diffractive events) as
well as Mueller-Navelet jets as a test of BFKL resummation, which can
be probed by measuring the difference in azimuthal angle between the
two most forward jets~\cite{Chevallier-talk}. The fraction of events
with gaps is about 10\% for soft double diffractive events and 1\% for
jet events with $E_T>2$ GeV in the miniplug
calorimeter~\cite{Mesropian-talk}. The $\Delta \eta$ dependence is
found to be similar for both samples.  It was reported
in~\cite{Roland-talk} that these measurements will also be possible at
the LHC thanks to the very good coverage of the forward region by the
ATLAS and CMS experiments.

\subsubsection{Inclusive diffraction at CDF}
\begin{figure}[t]
\begin{center}
\epsfig{file=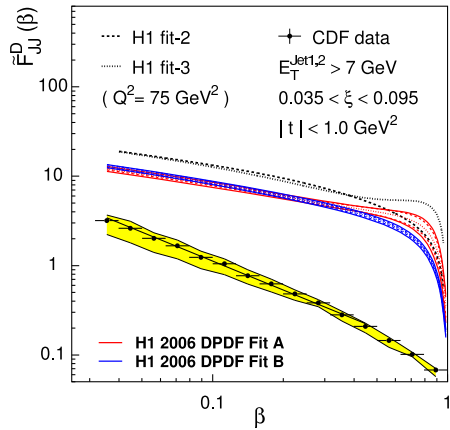,width=6cm}
\caption{Comparison between the CDF measurement of diffractive
  dijet production (black points) with the expectation from
  factorisation using H1 fits of diffractive PDFs (the blue band shows Fit
  B).} 
\label{Fig:fact}
\end{center}
\end{figure}

A long-standing issue in the field of diffraction has been the
complete breakdown of factorisation when diffractive PDFs from HERA
are used to predict measurements at the Tevatron
\cite{Newman-talk,Slominski-talk}.  The CDF collaboration measured the
ratio of dijet events in single diffractive and non-diffractive
events.  Assuming factorisation, this is directly proportional to the
ratio of the diffractive to the inclusive proton densities.  The CDF
measurement is compared with a factorised calculation using some of
the latest diffractive PDFs from H1 (Fit B and others) in
Fig.~$\ref{Fig:fact}$, where a clear discrepancy of an order of
magnitude can be seen.  Understanding this discrepancy at a
quantitative level is vital when making predictions for the LHC.

\subsubsection{Search for exclusive events at the Tevatron}

\begin{figure}
\begin{center}
\epsfig{file=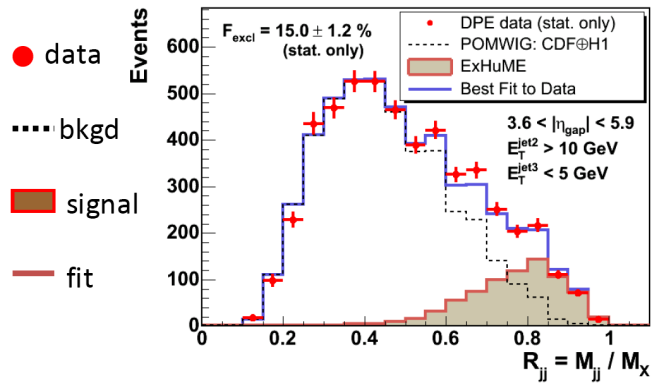,width=9cm}
\caption{Dijet mass fraction measured by the CDF collaboration compared to the
prediction adding the contributions from inclusive and exclusive diffraction.}
\label{exclusive}
\end{center}
\end{figure} 

Exclusive diffractive production, where the full energy is used to
produce the final state object (Higgs boson, dijets, diphotons...), is
a prominent topic in diffraction.  Such events benefit from forward
spectrometers with good energy resolution to provide a precise
measurement of the mass of the final state object.  CDF have looked for
the presence of such events in double-pomeron exchange events, where
the $\bar{p}$ is tagged and a rapidity gap is present on the other
side~\cite{Mesropian-talk}; the observable used is the
dijet mass fraction.  The comparison between the CDF data for a jet
$E_T$ cut of 10 GeV and the predictions from inclusive
diffraction is given in Fig.~\ref{exclusive}; the data can only be
described by the addition of a model of exclusive production.

Another interesting observable in the dijet channel is the rate of $b$
jets as a function of the dijet mass fraction. In exclusive events,
the $b$ jets are suppressed because of the $J_z=0$ selection rule, and
as expected, the fraction of $b$ jets in the diffractive dijet sample
diminishes as a function of the dijet mass
fraction~\cite{Mesropian-talk}.

Another way to look for exclusive events at the Tevatron is to search
for the diffractive exclusive production of light particles like the
$\chi$, $J/\Psi$ and $\Psi(2S)$ mesons~\cite{Mesropian-talk,
  Pinfold-talk}. Contrary to the diffractive exclusive production of
heavy mass objects such as Higgs bosons, this leads to high enough
cross sections to check the dynamical mechanisms and the existence of
exclusive events.  The CDF collaboration measured a $\chi$ production
cross section of $ \sim 75 \pm 14$ nb, compatible with theoretical
expectations.

The CDF collaboration also looked for the exclusive production of
diphotons~\cite{Mesropian-talk}.  Two exclusive diphoton events have been
observed by the CDF collaboration which is compatible with the
expectations for exclusive diphoton production at the Tevatron. An
update by the CDF collaboration with higher luminosity is
expected very soon.

\subsection{Factorisation and QCD fits}

The subject of rapidity gap survival probability can also be addressed
by the HERA experiments \cite{Newman-talk,Slominski-talk}.
Measurements of the ratio of diffractive dijets in photoproduction
with respect to diffractive dijets in DIS can be used to see if the
photon remnant, present in the resolved component of the
photoproduction events, produces a suppression.  Both H1 and ZEUS see
no difference in the ratio of cross sections between a
resolved-enriched sample of events and a direct-enriched sample.
However, the two experiments seem to disagree on the value of the
ratio itself, with H1 seeing clear evidence of a suppression of
photoproduction events, while ZEUS measures a ratio compatible with
there being no suppression.  The current status on this issue is that
both collaborations see evidence of an $E_T$ dependence (the
transverse energy of the leading jet) of the ratio; given the
different $E_T$ ranges of the two experiments, this could explain some
of the apparent discrepancy.  The normalisation uncertainty on the
ratio is of the order of 20\% and the ratio also depends on the choice
of diffractive PDFs used in the NLO prediction.  Further work towards
consistency in these respects would help to clarify the current
situation.

Another observable with potential sensitivity to rapidity gap
suppression is the ratio of diffractive to inclusive dijets.  This
ratio has now been measured by the H1 collaboration, reported in
\cite{Newman-talk}.  Unfortunately, the effect of the underlying
event, modelled by multiple interactions (MI) in PYTHIA, also affect
this ratio and are too large to be ignored.  The MI model describes
the data reasonably well, and the effect of including it or not yields
a difference in the ratio of approximately the same size as is
expected from suppression.

\begin{figure}[ht]
\begin{center}
\hspace{-1.0cm}
\includegraphics[width=0.15\textwidth]{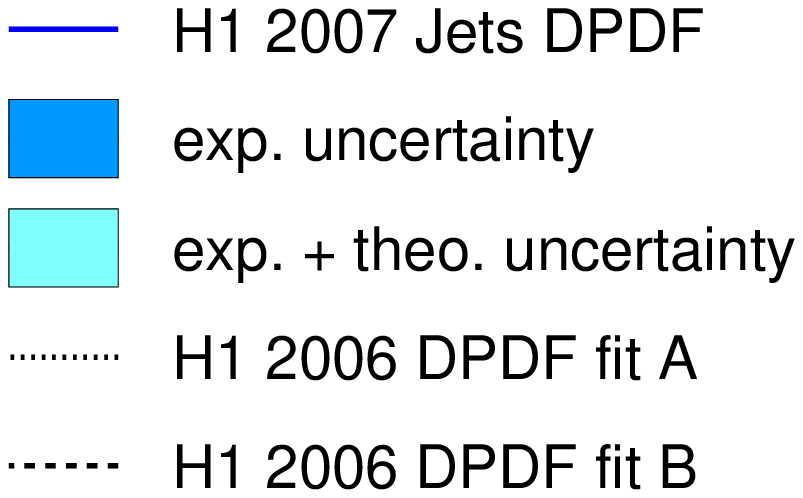}
\hspace{12.0cm}
\vspace{0.1cm}
\includegraphics[width=0.25\textwidth]{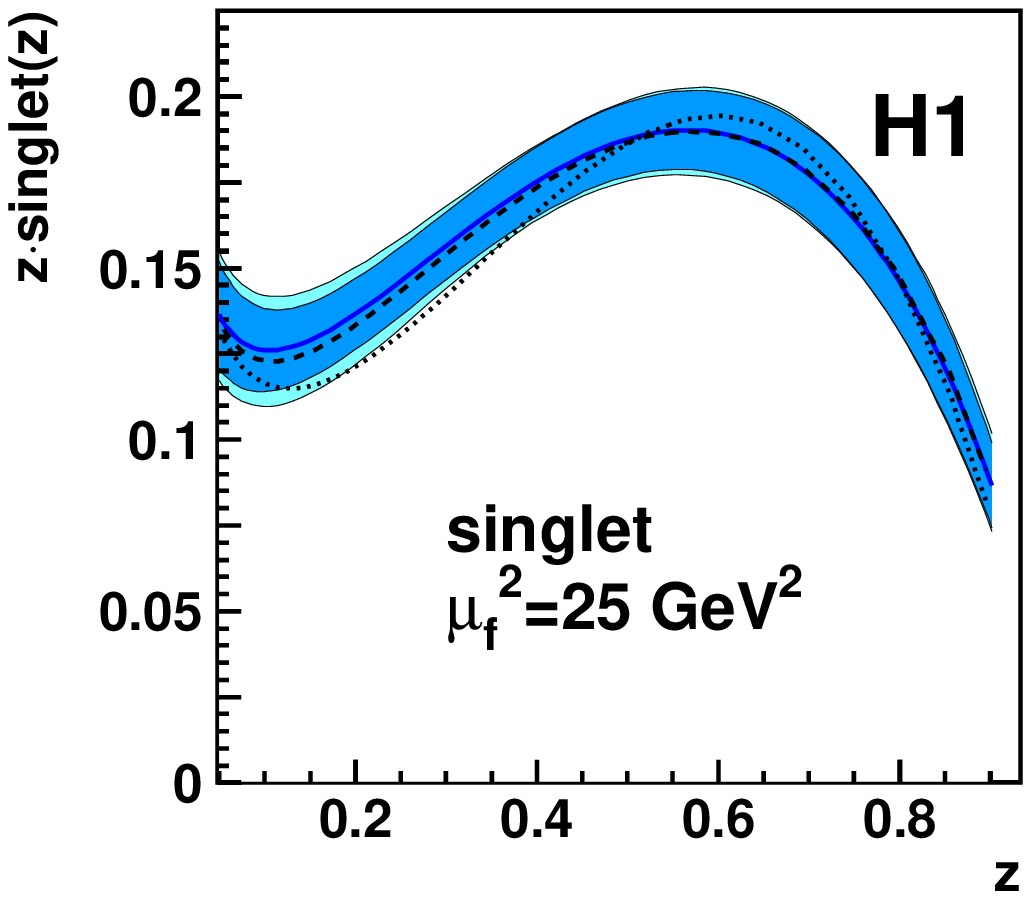}
\includegraphics[width=0.25\textwidth]{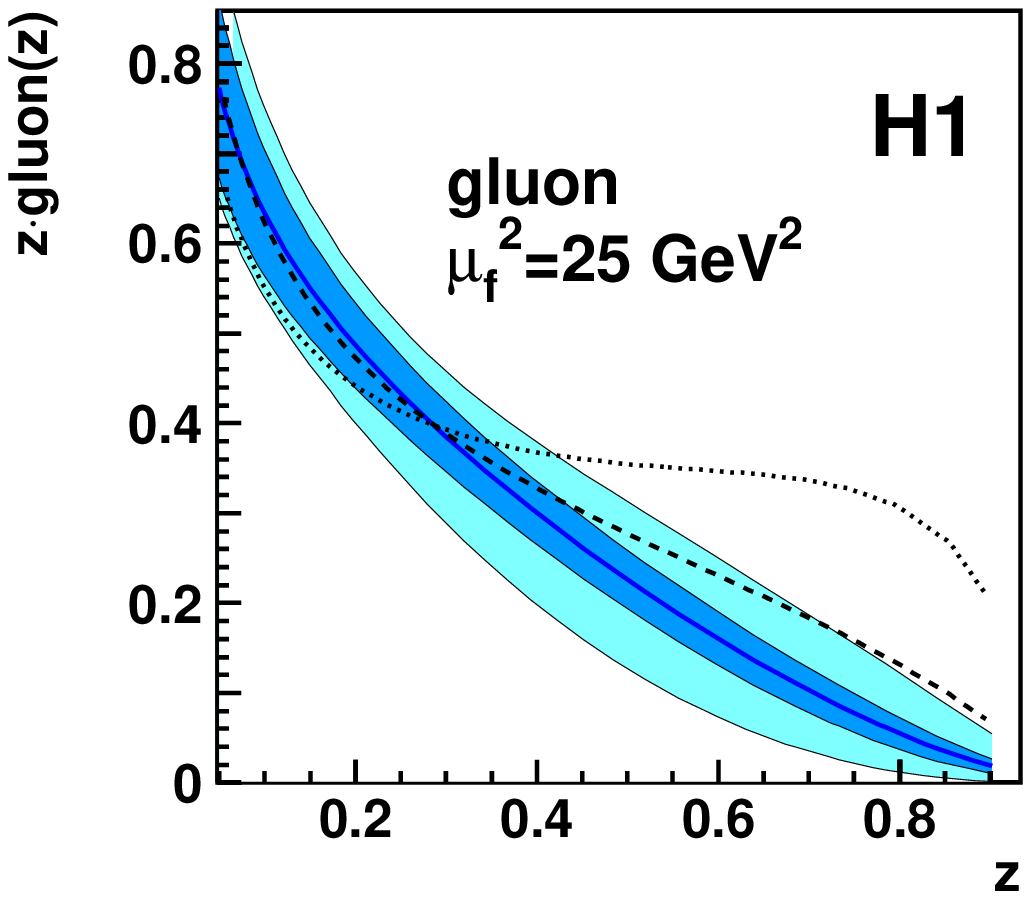}
\hspace{5.0cm}
\vspace{0.1cm}
\includegraphics[width=0.25\textwidth]{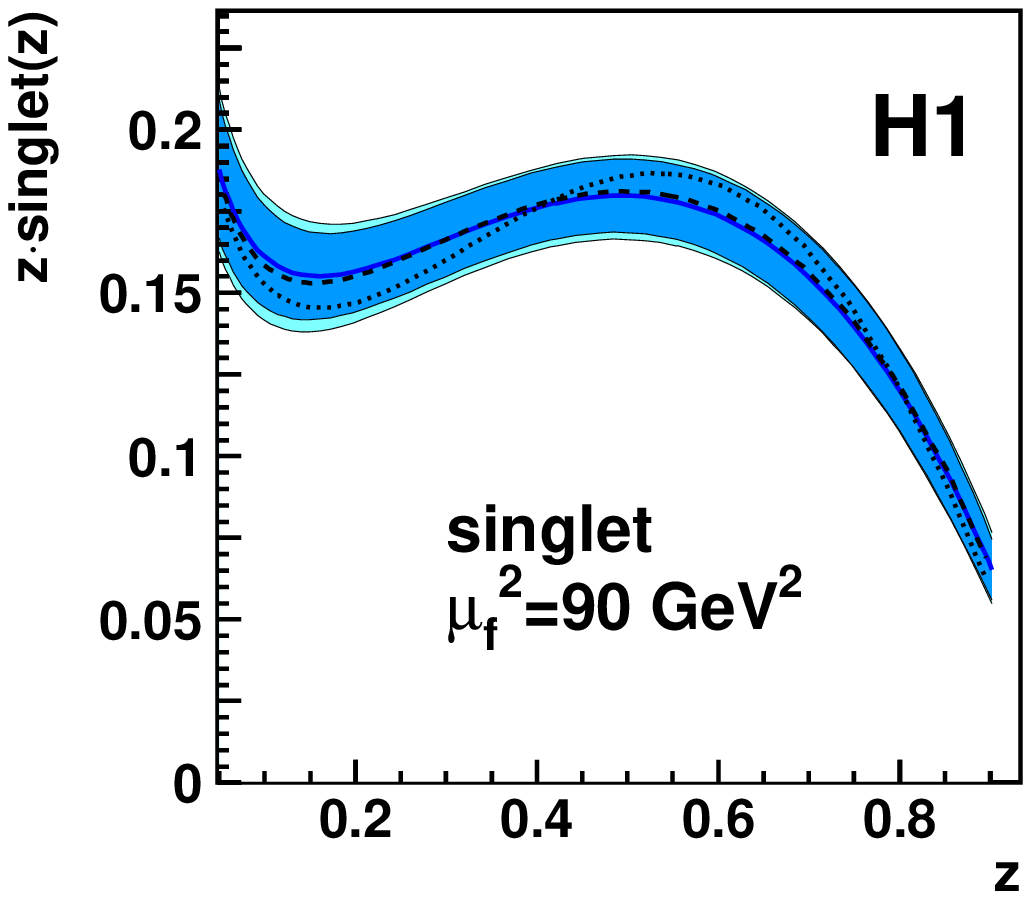}
\includegraphics[width=0.25\textwidth]{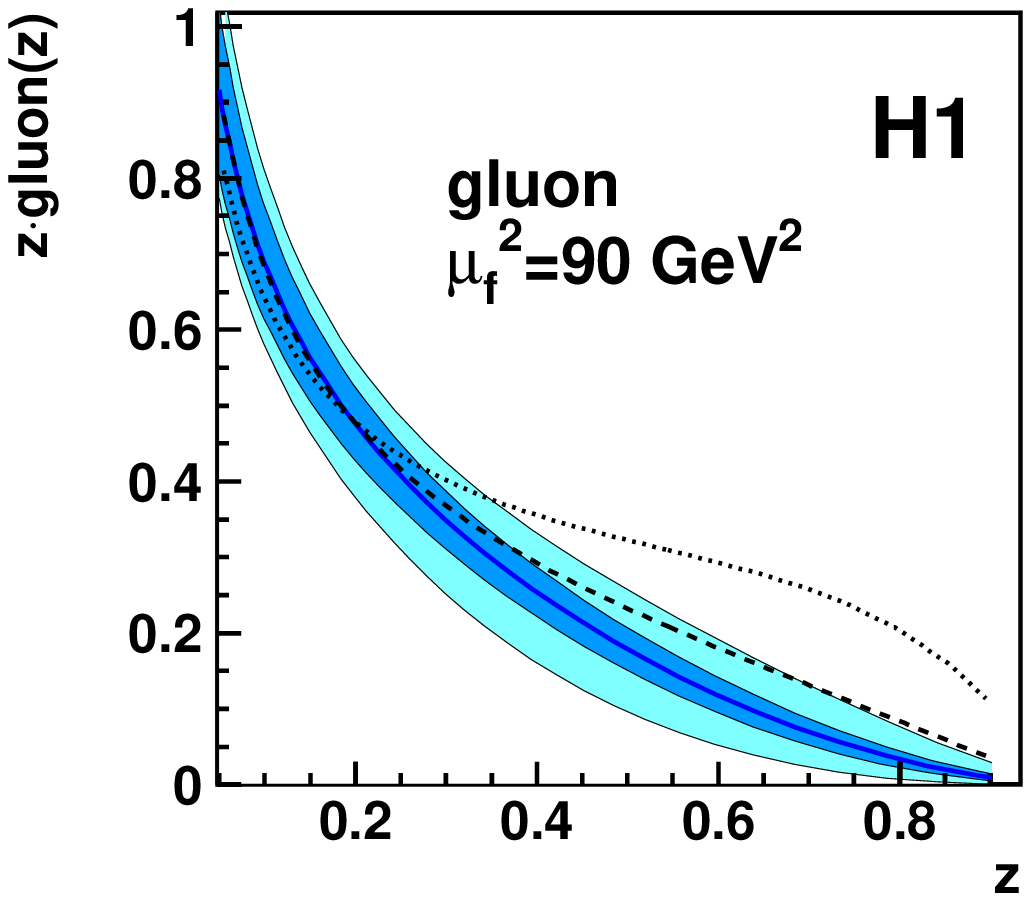}
\end{center}
\caption{The H1 diffractive PDFs resulting from 
the combined fit to the inclusive and dijet diffractive DIS data.}
\label{Fig:JetDPDFs}
\end{figure}

Despite the problems in understanding the photoproduction data, both
H1 and ZEUS have shown that the diffractive dijet data in DIS agree
very well with the factorisation assumption.  Both collaborations have
now made a combined fit to their inclusive and dijet DIS data, both
yielding results in very good agreement with one another, as was
reported in \cite{Slominski-talk}.  Fig. $\ref{Fig:JetDPDFs}$ shows a
summary plot of the latest published diffractive PDFs from the H1
Collaboration \cite{Aktas:2006hy}; the consistency between Fit B
(without dijet data) and the H1 2007 Jets PDF which include the dijet
data is good and gives us confidence that the NLO QCD picture used to
determine these distributions is indeed appropriate.

\subsection{Inclusive diffraction}

\begin{figure}[t]
\begin{center}
\epsfig{file=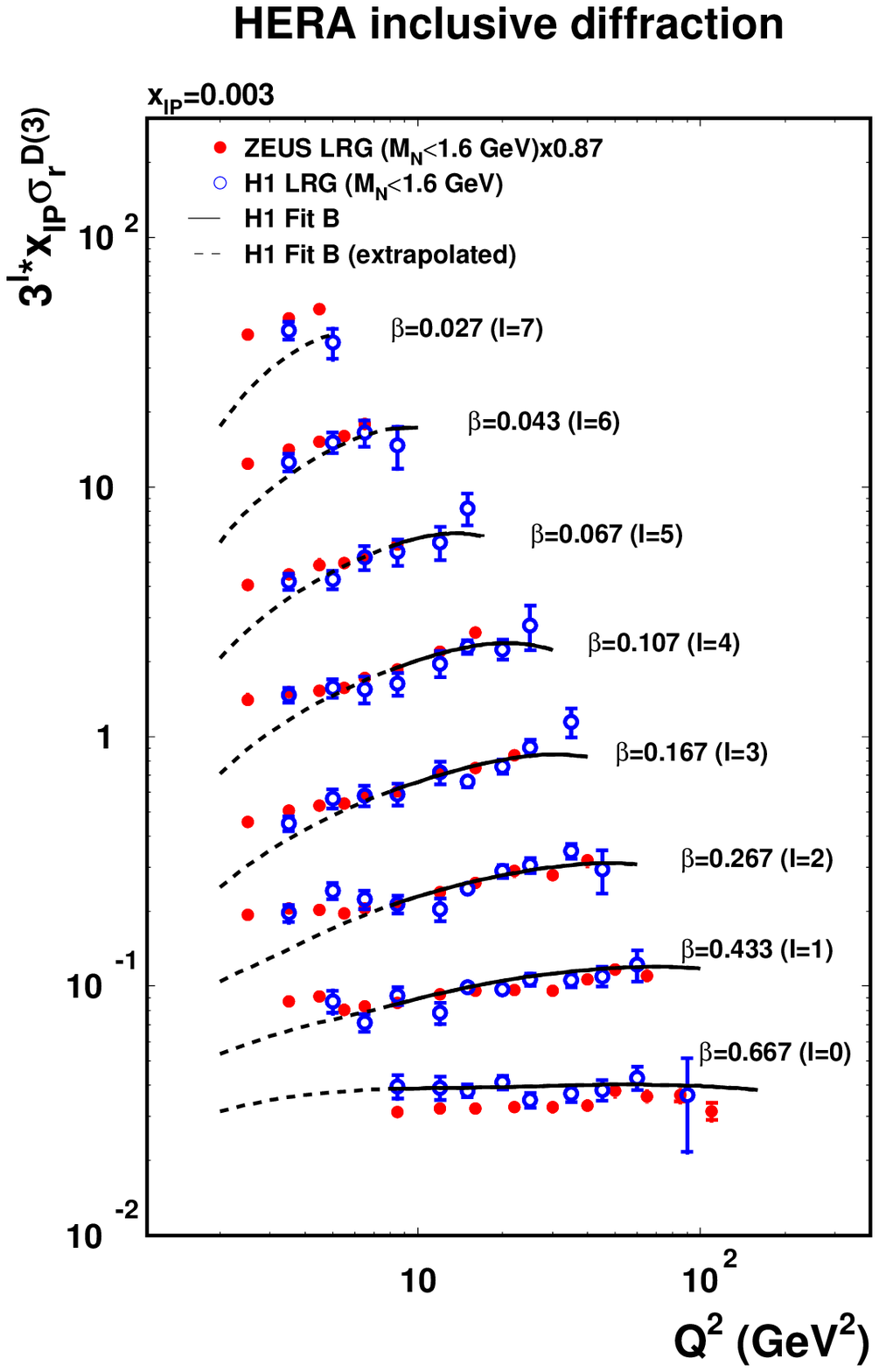,width=0.43\textwidth}
\epsfig{file=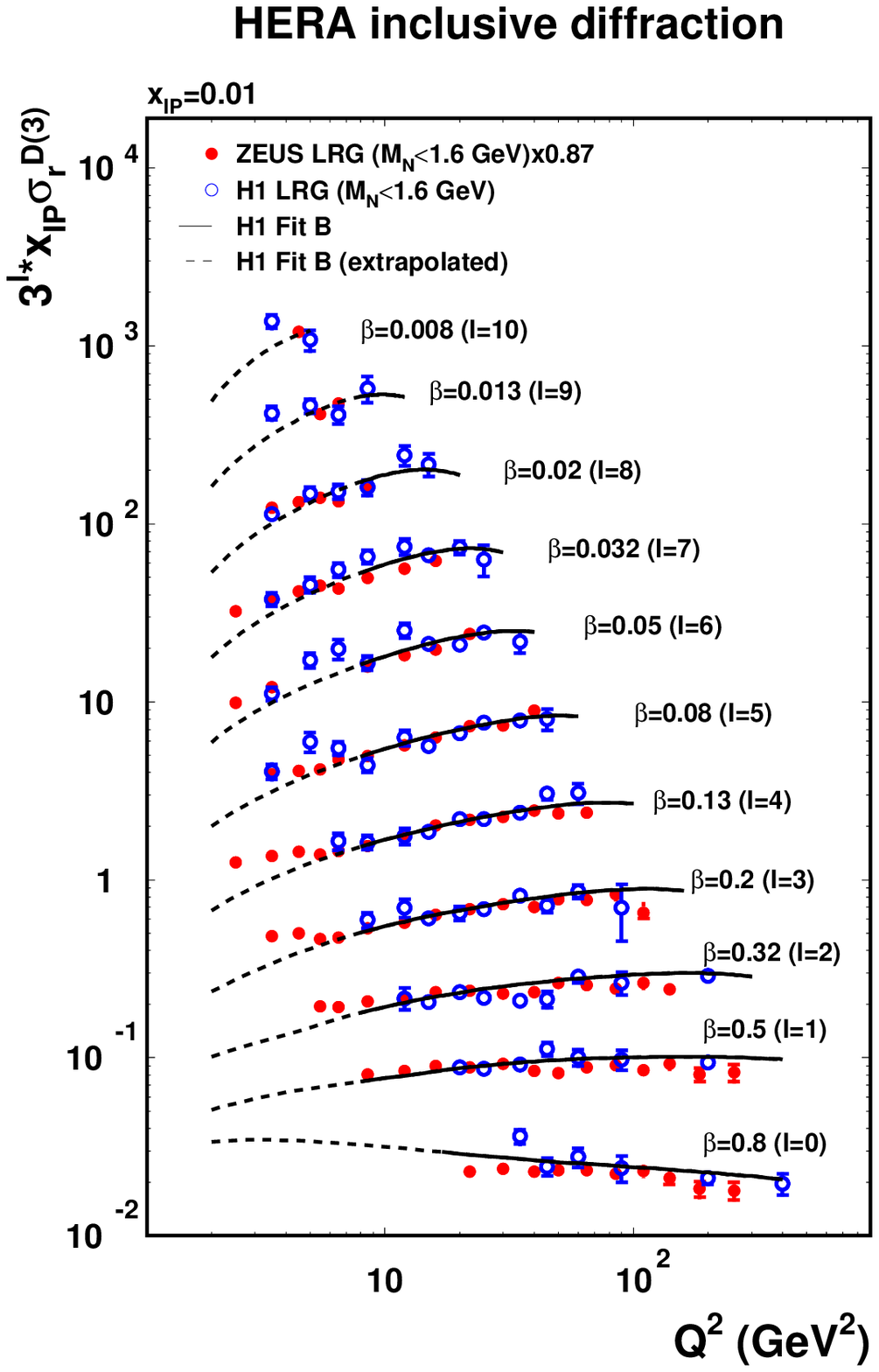,width=0.43\textwidth}
\caption{Comparison of the final ZEUS data on inclusive diffraction
  using the LRG method (red points) with the published H1 data (blue
  points) at two values of $x_{I\!\!P}$.}
\label{fig:H1ZeusIncComb}
\end{center}
\end{figure}

The final analysis of inclusive diffractive data from ZEUS
\cite{Chekanov:2008fh} using the 
large rapidity gap (LRG) method was reported in \cite{Ruspa-talk}.  The
data, discussed in 
terms of the reduced cross section, are shown in Fig.
$\ref{fig:H1ZeusIncComb}$ and compared to the published H1 LRG data
and to a QCD fit to that data (Fit B).  The precision of the ZEUS data
across such a broad kinematic range is very impressive and the final
measurements from H1 (using the full HERA II dataset) will be needed
to make a meaningful combination.

\subsection{First measurement of $F_L^D$}

The reduced cross section contains both diffractive structure
functions $F_2^D$ and $F_L^D$ and in most analyses is approximately
equal to $F_2^D$ due to the $y$-dependent suppression factor:
\begin{equation}
\sigma_r^D = F_2^D - \frac{y^2}{Y_{+}}F_L^D ,
\end{equation}
where $Y_{+}=1+(1-y)^2$.  The longitudinal diffractive structure
function, in analogy with the inclusive case, is approximately
proportional to the diffractive
gluon density, and a measurement of $F_L^D$ provides an independent
cross check of the gluon density extracted from the scaling violations
of $F_2^D$.  Experimentally, the longitudinal structure function can
be extracted from inclusive cross section measurements at different
beam energies.  Plotting $\sigma_r^D$ vs $y^2 /Y_{+}$ results
in a straight line with slope $-F_L^D$.

The final days of HERA running were devoted to two runs with proton
beam energy different from 920 GeV, first at 460 GeV and
finally at 575 GeV.  The measurement of $F_L^D$ using these data was
reported in \cite{Salek-talk}.  It is a technically challenging
measurement due to the need to measure at high values of $y$ (low
scattered electron energy), resulting in a large irreducible
photoproduction background that had to be understood in great
detail.

\begin{figure}
\begin{center}
\subfloat[\label{Fig:FLDXSecs}]{\epsfig{file=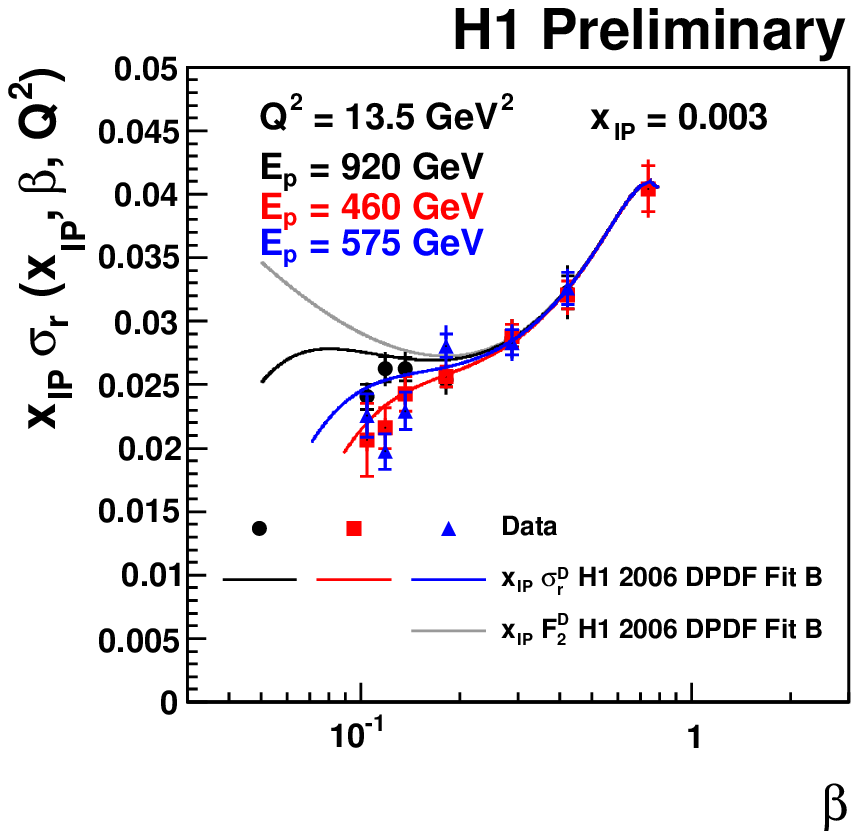,width=0.4\columnwidth}}
\subfloat[\label{Fig:FLD}]{\epsfig{file=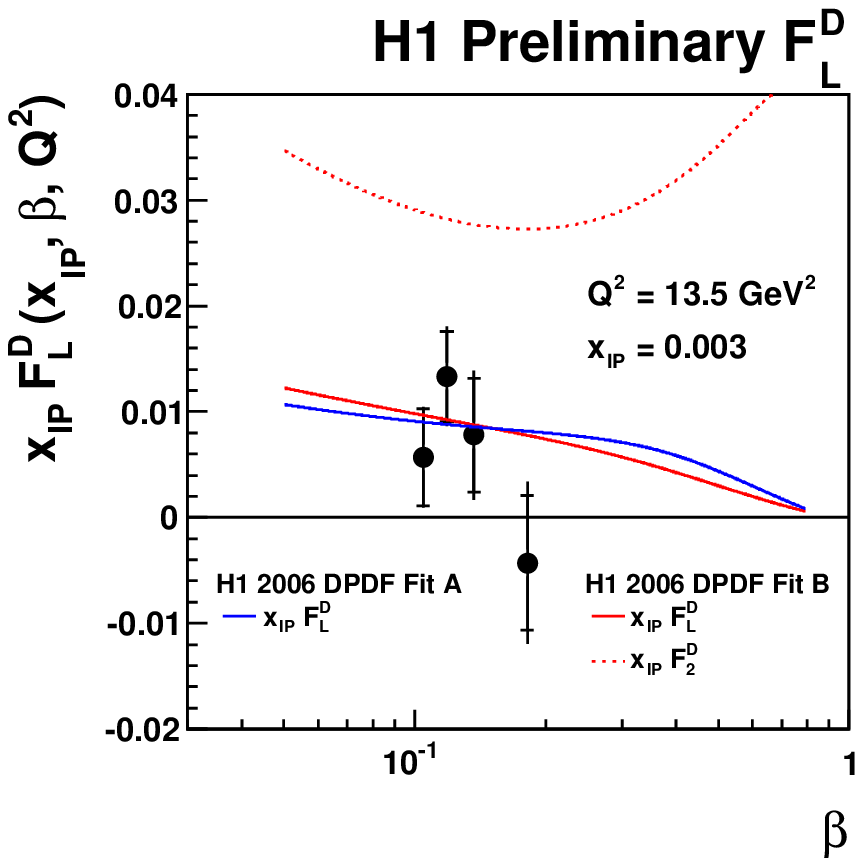,width=0.4\columnwidth}}
\caption{(a) The diffractive reduced cross section as a function of
  $\beta$ at three different beam energies.  (b) The diffractive
  longitudinal structure function $F_L^D$ as a function of $\beta$.}
\end{center}
\end{figure}

The diffractive reduced cross sections at each beam energy are shown
in Fig.~$\ref{Fig:FLDXSecs}$ as a function of $\beta$.  At high
$\beta$ (low $y$) the reduced cross section is equal to $F_2^D$ and
the data sets are normalised in this region to the result of H1 Fit B.
The effect of $F_L^D$ can be seen at the lowest $\beta$ values as a
suppression of the reduced cross section.  Fig.~$\ref{Fig:FLD}$
shows the measured value of $F_L^D$ compared to the prediction of Fit
B.  The result is a confirmation of the NLO QCD picture of diffraction
at greater than $3\sigma$.

\section{Theory}

\subsection{The colour dipole formalism}

A large number of theory presentations in the working group were
concerned with the colour dipole formalism.  This formalism
successfully describes a large variety of small-$x$ processes with a
\emph{common} non-perturbative input, namely the dipole scattering
amplitude.  Relevant observables are inclusive structure functions
($F_2$, $F_L$, $F_2^{c\bar{c}}$), diffractive structure functions
($F_2^{D}$, $F_2^{D (c\bar{c})}$), and exclusive diffractive channels
(vector meson production and DVCS).  The colour dipole formulation is
well suited to incorporate the dynamics of parton saturation, with the
dipole scattering amplitude entering in non-linear evolution equations
(JIMWLK, Balitsky-Kovchegov, and their generalisations).  Incorporated
into the larger framework of the colour glass condensate, the colour
dipole formulation provides a bridge between $ep$ scattering on one
hand and $pp$ and heavy-ion collisions on the other
\cite{Armesto-talk}.

\begin{figure}
\begin{center}
\includegraphics[width=0.75\textwidth]{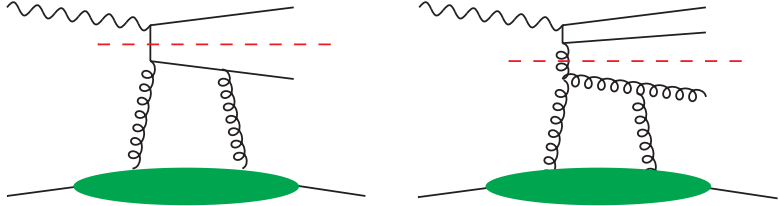}
\end{center}
\caption{\label{fig:dipole-graphs} Graphs for the production of $q\bar{q}$
  and $q\bar{q}g$ diffractive states in the colour dipole description.
  The horizontal dashed lines indicate the separation into a
  hard-scattering process and diffractive quark and gluon distributions in
  the large-$Q^2$ limit.}
\end{figure}

To a large extent, the description of inclusive diffraction in DIS is
based on two types of final states at parton level, as shown in
Fig.~\ref{fig:dipole-graphs}.  Complete theoretical expressions are
available for the $q\bar{q}$ diffractive final state.  It provides a
scaling contribution to $F_2^{D}$ and a $1/Q^2$ suppressed
contribution to $F_L^{D}$ at large $\beta$.  Its contribution to
$F_{2,L}^{D (c\bar{c})}$ is nonzero but numerically negligible in
typical HERA kinematics.  By contrast, available calculations for the
$q\bar{q}g$ final state are limited to either the leading $\log Q^2$
\cite{Luszczak-talk,Wusthoff:1997fz} or the leading $\log (1/\beta)$
\cite{Bartels:2002ri} approximation.  We note that in the latter case
one has strong ordering between the longitudinal momenta of the
final-state gluon and the $q\bar{q}$ pair, so that this configuration
plays a role in JIMWLK evolution and requires particular care when
applying this evolution to diffraction \cite{Weigert-talk}.  An
interpolation between the two logarithmic approximations has been
devised and used in phenomenology \cite{Lappi-talk}.  Unfortunately,
both approximate results have only been obtained for light quarks.
This is a rather unsatisfactory situation, because HERA measurements
have shown that the fraction of charm in the diffractive final state
can reach 20\% to 30\% \cite{Luszczak-talk}, well above what is
obtained from the $c\bar{c}$ final state in the dipole formulation.
Analyses omitting diffractive charm altogether are hence severely
limited in their precision, and it is difficult to assess the
reliability of the different workarounds so far used in dipole-based
descriptions.

A similar situation has emerged from the measurement of $F_L^{D}$
\cite{Salek-talk}, which has been one of the experimental highlights
in the diffractive session.  The data points cover the $\beta$-range
from 0.1 and 0.2, where the contribution from the $q\bar{q}$ final
state is completely negligible.  As for the $q\bar{q}g$ state, the
leading $\log (1/\beta)$ approximation is not applicable in this
region, whereas the leading $\log Q^2$ approximation gives a zero
result.\footnote{A $\log Q^2$ enhancement of the leading-twist
  contribution requires a nonzero leading-twist contribution at one
  order lower in $\alpha_s$, i.e.\ from the $q\bar{q}$ state.  Such a
  contribution is present for $F_T^{D}$, but not for $F_L^{D}$.}
This is clearly not good enough to describe the existing data for
$F_L^{D}$, and at the same time it may affect the accuracy for the
description of $F_2^{D} = F_T^{D} + F_L^{D}$.

By now, analyses of HERA data within the dipole formulation typically
address subtle questions such as the importance of parton saturation or
the dependence on the impact parameter and its interplay with other
kinematic variables.  This requires sufficient precision of the theory,
and in this respect the current data call for an improved theoretical
evaluation of the $q\bar{q}g$ final state, for both light and heavy
quarks.

It has been realised long ago \cite{Buchmuller:1998jv} that the dipole
formulation for $q\bar{q}$ and $q\bar{q}g$ final states can be used to
obtain a simple prediction for diffractive PDFs, using suitable
approximations of the dipole results in kinematics where diffractive PDFs
can be used.  The resulting PDFs pertain to a low scale; DGLAP evolution
then accounts for further radiation leading to final states $q\bar{q}gg$
etc.  This idea has been revived in a recent study \cite{Luszczak-talk},
where the diffractive PDFs extracted from a dipole model were compared
with those obtained in a DGLAP fit to $F_2^{D}$.  Quite good agreement was
found for the diffractive quark distribution at momentum fractions $z >
0.3$, where the generation of low-momentum sea quarks by evolution does
not play a major role.  In turn, a clear discrepancy was observed for the
diffractive gluon density $g^D(z)$, with the distribution extracted from
the dipole model vanishing smoothly for $z\to 1$, whereas the DGLAP fit
gave a pronounced peak at high $z$ for moderate factorisation scales.

This should, however, not be regarded as a shortcoming of the dipole
result.  Confirming earlier findings, a recent dedicated analysis
\cite{Schoeffel-talk} has emphasised that $F_2^{D}$ data alone provide
only a poor constraint on $g^D(z)$ at high $z$.  (An important result of
the same study is that even within the resulting uncertainty, the
diffractive gluon density is not sufficiently large at high $z$ to account
for the rate of exclusive diffractive dijets observed by CDF
\cite{Mesropian-talk}.  This underlines the need for a genuinely exclusive
production mechanism.)  Furthermore, the data for diffractive dijet
production clearly disfavour a pronounced bump of $g^D(z)$ at high $z$
compared with a smooth drop to zero \cite{Newman-talk}.  {}From this
perspective, the dipole model result for $g^D(z)$ in \cite{Luszczak-talk}
seems to be in line with DGLAP phenomenology.

An important property of the dipole scattering amplitude is its dependence
on the impact parameter $b$.  It describes the spatial distribution of
colour charge in the target and is of particular importance in the context
of saturation, which will set in earlier in the dense centre of a hadron
than at its more dilute edge.  The $t$ dependence of inclusive and
exclusive diffractive cross sections measured at HERA plays an essential
role in attempts to determine the $b$ dependence of the dipole scattering
amplitude.  A non-trivial pattern of preferred $b$ values in diffractive
$q\bar{q}$ and $q\bar{q}g$ states has been discussed in \cite{Lappi-talk},
and characteristic predictions for inclusive diffraction on nuclei have
been deduced, with a suppression at low and an enhancement at high $\beta$
compared to the proton case.  As to exclusive diffraction on nuclei,
detailed studies for vector meson production and DVCS have been given in
\cite{Machado-talk,Marquet-talk}, with emphasis on the differences between
the elastic contribution and nuclear breakup.

\subsection{Generalised parton distributions and exclusive diffraction}

The concept of generalised parton distributions (GPDs) permits the
description of exclusive processes in a wide kinematic regime, and it
offers the possibility of three-dimensional imaging of partons in the
nucleon or in nuclei.  An impressive amount of relevant data has been
taken both in fixed-target experiments \cite{Spin-summary}
and in the small-$x$ regime of H1 and ZEUS \cite{Levy-talk,Marage-talk}.
There are encouraging prospects of measuring exclusive $J/\Psi$ and
$\Upsilon$ photoproduction on the proton or nuclei at the LHC
\cite{Csanad-talk}, which would extend existing measurements of these
theoretically rather clean channels to very high energies.  Additionally,
ultraperipheral collisions at LHC offer the opportunity to study timelike
Compton scattering, $\gamma p\to \ell^+\ell^- p$ \cite{Wagner-talk}.
Comparison of this reaction with DVCS at small $x$ measured at HERA would
permit a key test for the theory description, which is closely related for
the underlying QCD processes $\gamma^* p\to \gamma p$ and $\gamma p\to
\gamma^* p$ (with a spacelike photon in one and a timelike photon in the
other case).  Finally, generalised parton distributions are an important
input to the description of exclusive diffraction in $pp$ collisions at
the LHC \cite{Weiss-talk}.

The clarification of a long-standing theoretical dispute has been reported
in the working group.  In an early paper, Shuvaev et
al.~\cite{Shuvaev:1999ce} claimed that in the small-$x$ limit, GPDs at
$t=0$ can be \emph{calculated} from the usual parton densities.  A simple
formula was derived for the case where the PDFs follow a power-law in $x$
and has been used in many phenomenological analyses.  It is now clear
\cite{Muller-talk,Teubner-talk} that this claim cannot be upheld.  The
original derivation in \cite{Shuvaev:1999ce} fails due to subtleties in an
inverse moment transform (which generalises the inverse Mellin transform
for conventional PDFs), and there exist explicit examples of GPDs that
satisfy all known symmetry relations but contradict the Shuvaev formula.
Additional properties are required to ensure the validity of this formula,
which given our present understanding should be understood as an
\emph{ansatz}, to be checked against data.  A detailed study of this issue
for both quark and gluon distributions has been reported in
\cite{Muller-talk}.  At the same time, the Shuvaev ansatz can readily be
used for PDFs that do not exhibit a power behavior, and a corresponding
numerical code is available \cite{Teubner-talk}.

Exclusive Higgs production has long been considered a highlight for
diffractive physics at the LHC.  Its potential to explore physics beyond
the standard model has been reviewed in \cite{Khoze-talk}, with emphasis
on the minimal supersymmetric Standard Model.  On the strong-interaction
side, a major open issue remains the evaluation of the rapidity-gap
survival probability for exclusive diffraction in LHC kinematics; a topic
that merits additional attention due to its connection with the physics of
multiple parton interactions \cite{Bartels-talk}.  The assessment
presented in \cite{Weiss-talk} is that interactions involving \emph{hard}
spectator partons may be more important than assumed in most present
estimates.  This effect would increase with energy and could decrease the
survival probability at the LHC without affecting the corresponding
estimates for Tevatron kinematics, which are consistent with the observed
rate of exclusive diffractive dijets \cite{Mesropian-talk}.  More detailed
theoretical work will be needed to clarify this issue.

\subsection{Further theoretical developments}

Theoretical progress on many more aspects of diffraction has been reported
in the working group and can only be briefly mentioned in this summary.

Work on a unified description of diffraction with either an intact
final-state proton or with proton dissociation was reported in
\cite{Marquet-talk}, with the aim of treating diffractive measurements
at low and high $t$ in a common formalism based on the colour glass
condensate.  An analysis of Mueller-Navelet jets and rapidity gaps
between jets in the BFKL framework was presented in
\cite{Chevallier-talk}.  NLO corrections to the BFKL kernel were found
to be moderate, and a comparison with Tevatron data was given, as well
as predictions for the LHC.

The exclusive production of forward jets, $pp \to p + 3 ~\text{jets}$
was explored in the $k_T$ factorisation framework \cite{Ivanov-talk}.
This process is sensitive to the $k_T$ unintegrated gluon GPD in the
proton on one side and the three-quark proton distribution amplitudes
on the other, and it will be interesting to see whether it can be
measured with forward detectors at the LHC.  The process $\gamma\gamma
\to (\pi^+\pi^-) + (\pi^+\pi^-)$ was investigated in
\cite{Schwennsen-talk}, where it was shown that appropriate charge
asymmetries in the final state project out the interference between
perturbative odderon and pomeron exchange.  Estimates indicate that
the asymmetry may be large, but more detailed experimental studies
will be required to establish whether this channel can be observed in
ultraperipheral $pp$ or heavy-ion collisions at the LHC.

Various extensions of the Standard Model predict new heavy
strongly-interacting particles.  Some of these would be metastable and
hence form bound states with ordinary quarks or gluons, before
decaying inside the LHC detectors.  The interactions of such heavy
hadrons with the detector material were studied in
\cite{Milstead-talk}.  An important finding was that in certain cases
an initially charged hadron tends to be converted to a neutral one by
repeated scattering, which is of immediate consequence for
experimental search strategies.

\section{Conclusions}

Although in general we see that the gross features are understood in
QCD, there remain many questions to be answered in the fields of
diffractive and vector meson physics.  The issue of rapidity gap
survival probability may be qualitatively understood, but the
experimental status of the subject at HERA remains less than ideal.
Nevertheless, the experimental results from the Tevatron on exclusive
production agree well with theory and bode well for extrapolating our
current understanding to the LHC domain.

The HERA data have helped to elucidate, confirm and reject much; the
measurement of $F_L^D$ confirms that NLO QCD is applicable to
inclusive diffraction in DIS.  The exclusive vector meson and DVCS
results can be understood in terms of models based on either GPDs or
the colour dipole approach.  Nevertheless, precision can still be
improved here, and the final publications from H1 and ZEUS as well as
data combinations are eagerly anticipated and needed.  These, together
with the promising physics program of the VFPS of H1, mean that the
final word from HERA has not yet been uttered.

\section{Acknowledgments}

We would like to thank all speakers in the working group for their
carefully prepared contributions, our co-convenors from the working groups
on Structure Functions and low-$x$ and on Spin Physics for organising
common sessions, and J.~Terron, C.~Glasman, A.~Sabio Vera and
C.~Uribe-Estrada for hosting a wonderful meeting.

\section{Bibliography}

\begin{footnotesize}

\end{footnotesize}

\end{document}